\documentclass[aps,prb,twocolumn,twoside,superscriptaddress,longbibliography]{revtex4-2}

\usepackage[version=3]{mhchem} 
\usepackage{graphicx}
\usepackage{amsmath}
\usepackage{float}
\usepackage{amssymb}
\usepackage[dvipsnames]{xcolor}
\usepackage{changes}
\usepackage{mathrsfs}
\usepackage{comment}
\usepackage{multirow}

\begin{document}

\title{Substrate Effects on Spin Relaxation in Two-Dimensional Dirac Materials with Strong Spin-Orbit Coupling} 
\setcounter{page}{1} 
\date{\today}

\author{Junqing Xu}\email{jxu153@ucsc.edu} 
\affiliation{Department of Chemistry and Biochemistry, University of California, Santa Cruz, CA 95064, USA}
\author{Yuan Ping}\email{yuanping@ucsc.edu} 
\affiliation{Department of Chemistry and Biochemistry, University of California, Santa Cruz, CA 95064, USA}
\begin{abstract}
Understanding substrate effects on spin dynamics and relaxation in
two-dimensional (2D) materials is of key importance for spintronics
and quantum information applications. However, the key factors that
determine the substrate effect on spin relaxation, in particular for
materials with strong spin-orbit coupling, have not been well understood.
Here we performed first-principles real-time density-matrix dynamics
simulations with spin-orbit coupling (SOC) and quantum descriptions
of electron-phonon and electron-impurity scattering for the spin lifetimes
of supported/free-standing germanene, a prototypical strong SOC 2D
Dirac material. We show that the effects of different substrates on
spin lifetime ($\tau_{s}$) can surprisingly differ by two orders
of magnitude. We find that substrate effects on $\tau_{s}$
are closely related to substrate-induced modifications of the SOC-field
anisotropy, which changes the spin-flip scattering matrix elements.
We propose a new electronic quantity, named spin-flip angle $\theta^{\uparrow\downarrow}$, to characterize spin relaxation caused by intervalley
spin-flip scattering. We find that the spin relaxation rate
is approximately proportional to the averaged value of $\mathrm{sin}^{2}\left(\theta^{\uparrow\downarrow}/2\right)$, which can be used as a guiding parameter of controlling spin relaxation.

\end{abstract}
\maketitle

\section*{Introduction}

Since the long spin diffusion length ($l_{s}$) in large-area graphene
was first reported by Tombros et al.\citep{tombros2007electronic},
significant advances have been made in the field of spintronics, which
has the potential to realize low-power electronics by utilizing spin
as the information carrier. Various 2D materials have shown promising
spintronic properties\citep{avsar2020colloquium}, e.g., long $l_{s}$
at room temperatures in graphene\citep{drogeler2016spin} and ultrathin
black phosphorus\citep{avsar2017gate}, spin-valley locking (SVL)
and ultralong spin lifetime $\tau_{s}$ at low temperatures in transition
metal dichalcogenides (TMDs)\citep{dey2017gate} and germanene\citep{xu2021giant}, and 
persistent spin helix in 2D hybrid perovskites\citep{zhang2022room}.

Understanding spin relaxation and transport mechanism in materials
is of key importance for spintronics and spin-based quantum information
technologies. One critical metric for ideal materials in such applications
is spin lifetime ($\tau_{s}$), often required to be sufficiently
long for stable detection and manipulation of spin. In 2D-material-based
spintronic devices, the materials are usually supported on a substrate.
Therefore, for the design of those devices, it is crucial to understand
substrate effects on spin relaxation. In past work, the substrate
effects were mostly studied for weak SOC Dirac materials like graphene\cite{habib2022electric,ertler2009electron,cummings2017giantspinlifetime,van2016spin,zhang2012electron}.
How substrates affect strong SOC Dirac materials like germanene is
unknown. In particular, the spin relaxation mechanism between weak
and strong SOC Dirac materials was shown to be drastically different.~\citep{xu2021giant}
Therefore, careful investigations are required to unveil the distinct
substrate effects on these two types of materials.

Here we focus on the dangling-bond-free insulating substrates, which
interact weakly with the material thus preserve its main physical
properties. Insulating substrates can affect spin dynamics and relaxation
in several aspects: (i) They may induce strong SOC fields, so called
internal magnetic fields ${\bf B}^{\mathrm{in}}$ by breaking inversion
symmetry\citep{ertler2009electron} or through proximity effects\citep{cummings2017giantspinlifetime}.
For example, the hexagonal boron nitride substrate can induce Rashba-like
fields on graphene and dramatically accelerate its spin relaxation
and enhance the anisotropy of $\tau_{s}$ between in-plane and out-of-plane
directions\citep{habib2022electric}. (ii) Substrates may introduce
additional impurities \citep{van2016spin,zhang2012electron} or reduce
impurities/defects in material layers, e.g., by encapsulation\citep{li2021valley}.
In consequence, substrates may change the electron-impurity (e-i)
scattering strength, which affects spin relaxation through SOC. 
(iii) Thermal vibrations of substrate atoms can introduce additional
spin-phonon scattering by interacting with spins of materials\citep{ertler2009electron}.

Previously most theoretical studies of substrate effects on spin relaxation
were done based on model Hamiltonian and simplified spin relaxation
models\citep{ertler2009electron,van2016spin,zhang2012electron}. While
those models provide rich mechanistic insights, they are lack of predictive
power and quantitative accuracy, compared to first-principles theory.
On the other hand, most first-principles studies only simulated the
band structures and spin polarizations/textures of the heterostructures\citep{ni2015electronic,amlaki2016z,zollner2021graphene},
which are not adequate for understanding spin relaxation. Recently,
with our newly-developed first-principles density-matrix (FPDM) dynamics
approach, we studied the hBN substrate effect on spin relaxation of
graphene, a weak SOC Dirac material. We found a dominant D'yakonov-Perel'
(DP) mechanism and nontrivial modification of SOC fields and electron-phonon
coupling by substrates\cite{habib2022electric}. However, strong SOC
Dirac materials can have a different spin relaxation mechanism - Elliott-Yafet
(EY) mechanism\citep{vzutic2004spintronics}, with only spin-flip
transition and no spin precession, unlike the DP mechanism. How substrates
affect spin relaxation of materials dominated by EY mechanism is the
key question here. Furthermore, how such effects vary among different
substrates is another outstanding question for guiding experimental
design of interfaces.

In our recent study, we have predicted that monolayer germanene (ML-Ge)
is a promising material for spin-valleytronic applications, due to
its excellent properties including spin-valley locking, long $\tau_{s}$
and $l_{s}$, and highly tunable spin properties by varying gates
and external fields\citep{xu2021giant}. As discussed in Ref.~\citenum{xu2021giant},
ML-Ge has strong intrinsic SOC unlike graphene and silicene. Under
an out-of-plane electric field (in consequence broken inversion symmetry), a strong
out-of-plane internal magnetic field forms, which may lead to mostly
EY spin relaxation~\cite{xu2021giant}. Therefore, predicting $\tau_{s}$
of supported ML-Ge is important for future applications and our understanding
of substrate effects on strong SOC materials. 
Here, we examine the substrate effects on spin relaxation in ML-Ge
through FPDM simulations, with self-consistent SOC and quantum descriptions
of e-ph and e-i scattering processes\citep{xu2020spin,xu2021ab,xu2021giant,habib2022electric,xu2022spin}.
We study free-standing ML-Ge and ML-Ge supported by four different
insulating substrates - germanane (GeH), silicane (SiH), GaTe and
InSe. The choice of substrates is based on similar lattice constants
to ML-Ge, preservation of Dirac Cones, and experimental synthesis
accessibility\citep{giousis2021synthesis,lei2014evolution}. We will
first show how electronic structures and $\tau_{s}$ of ML-Ge are
changed by different substrates - while $\tau_{s}$
of ML-Ge on GeH and SiH are similar to free-standing ML-Ge, the GaTe
and InSe substrates strongly reduce $\tau_{s}$ of ML-Ge due to stronger
interlayer interactions. We then discuss what quantities are responsible
for the disparate substrate effects on spin relaxation, which eventually
answered the outstanding questions we raised earlier.

\section*{Results and discussions}

\subsection*{Substrate effects on electronic structure and spin texture}

\begin{figure*}[!ht]
\includegraphics[scale=0.54]{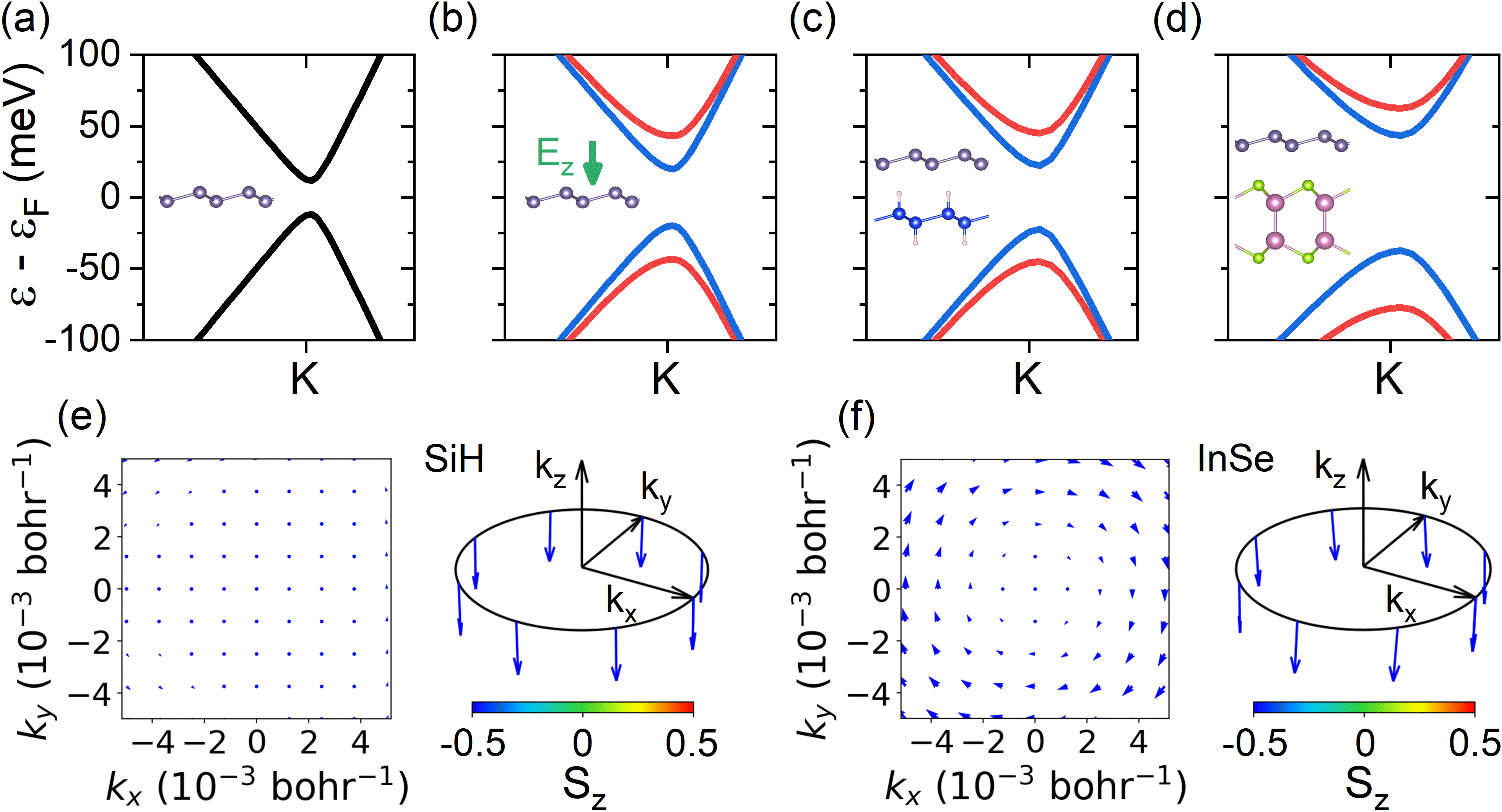}

\caption{Band structures and spin textures around the Dirac
cones of ML-Ge systems with and without substrates. (a)-(d) show
band structures of ML-Ge under $E_{z}=0$ and under -7 V/nm and ML-Ge
on silicane (SiH) and on InSe substrates respectively.
(e) and (f) show spin textures in the $k_{x}$-$k_{y}$ plane and
3D plots of the spin vectors ${\bf S}_{k_{1}}^{\mathrm{exp}}$ on
the circle $|\protect\overrightarrow{k}|=0.005$ bohr$^{-1}$ of the
band at the band edge around $K$ of ML-Ge on SiH and InSe substrates
respectively. ${\bf S}^{\mathrm{exp}}\equiv\left(S_{x}^{\mathrm{exp}},S_{y}^{\mathrm{exp}},S_{z}^{\mathrm{exp}}\right)$
with $S_{i}^{\mathrm{exp}}$ being spin expectation value along direction
$i$ and is the diagonal element of spin matrix $s_{i}$ in Bloch
basis. The red and blue bands correspond to spin-up and spin-down
states. Due to time-reversal symmetry, band structures around another
Dirac cone at $K'=-K$ are the same except that the spin-up and spin-down
bands are reversed. The grey, white, blue, pink and green balls correspond
to Ge, H, Si, In and Se atoms, respectively. Band structures of ML-Ge
on germanane (GeH) and GaTe are shown in Fig. S4 in the Supporting
Information, and are similar to those of ML-Ge on SiH and InSe substrates,
respectively. In subplots (e) and (f), the color
scales $S_{z}^{\mathrm{exp}}$ and the arrow length scales the vector
length of in-plane spin expectation value.\label{fig:bands}}
\end{figure*}

We begin with comparing band structures and spin textures of free-standing
and supported ML-Ge in Fig.~\ref{fig:bands}, which are essential
for understanding spin relaxation mechanisms. 
Since one of the most important effects of a substrate is to induce an out-of-plane
electric field $E_{z}$ on the material layer, we also study ML-Ge
under a constant $E_{z}$ as a reference. 
The choice of the $E_{z}$ is based on
reproducing a similar band splitting to the one in ML-Ge with substrates.
The band structure of ML-Ge is similar to graphene with two Dirac
cones at $K$ and $K'\equiv-K$, but a larger band gap of 23 meV. At $E_{z}=0$,
due to time-reversal and inversion symmetries of ML-Ge, every two
bands form a Kramers degenerate pair\citep{vzutic2004spintronics}.
A finite $E_{z}$ or a substrate breaks the inversion symmetry and
induces a strong out-of-plane internal B field ${\bf B}^{\mathrm{in}}$
(Eq.~\ref{eq:Bin}), which splits the Kramers pairs into spin-up and
spin-down bands\citep{xu2021giant}. Interestingly, we find that band
structures of ML-Ge-SiH (Fig.~\ref{fig:bands}c) and ML-Ge-GeH (Fig.
S4) are quite similar to free-standing ML-Ge under $E_{z}$=-7 V/nm
(ML-Ge@-7V/nm, Fig.~\ref{fig:bands}b), which indicates that the
impact of the SiH/GeH substrate on band structure and ${\bf B}^{\mathrm{in}}$
may be similar to a finite $E_{z}$ (see Fig. S4). This similarity
is frequently assumed in model Hamiltonian studies\citep{van2016spin,ertler2009electron}.
On the other hand, the band structures of ML-Ge-InSe (Fig.~\ref{fig:bands}d)
and ML-Ge-GaTe (Fig. S4) have more differences from the free-standing
one under $E_{z}$, with larger band gaps, smaller band curvatures
at Dirac Cones, and larger electron-hole asymmetry of band splittings.
This implies that the impact of the InSe/GaTe substrates can not be
approximated by applying an $E_{z}$ to the free-standing ML-Ge, unlike
SiH/GeH substrates.

We further examine the spin expectation value vectors
${\bf S}^{\mathrm{exp}}$ of substrate-supported ML-Ge. ${\bf S}^{\mathrm{exp}}$ is parallel to ${\bf B}^{\mathrm{in}}$ by definition (Eq. \ref{eq:Bin}). ${\bf S}^{\mathrm{exp}}\equiv\left(S_{x}^{\mathrm{exp}},S_{y}^{\mathrm{exp}},S_{z}^{\mathrm{exp}}\right)$
with $S_{i}^{\mathrm{exp}}$ being spin expectation value along direction
$i$ and is the diagonal element of spin matrix $s_{i}$ in Bloch
basis. Importantly, from Fig.~\ref{fig:bands}e and \ref{fig:bands}f,
although ${\bf S}^{\mathrm{exp}}$ of ML-Ge on substrates are highly
polarized along $z$ (out-of-plane) direction, the in-plane components of ${\bf S}^{\mathrm{exp}}$
of ML-Ge-InSe (and ML-Ge-GaTe) are much more pronounced than ML-Ge-SiH
(and ML-Ge-GeH). Such differences are crucial to the out-of-plane
spin relaxation as discussed in a later subsection.

\subsection*{Spin lifetimes of germanene on substrates and spin relaxation mechanism}
\begin{figure*}[!ht]
\includegraphics[scale=0.78]{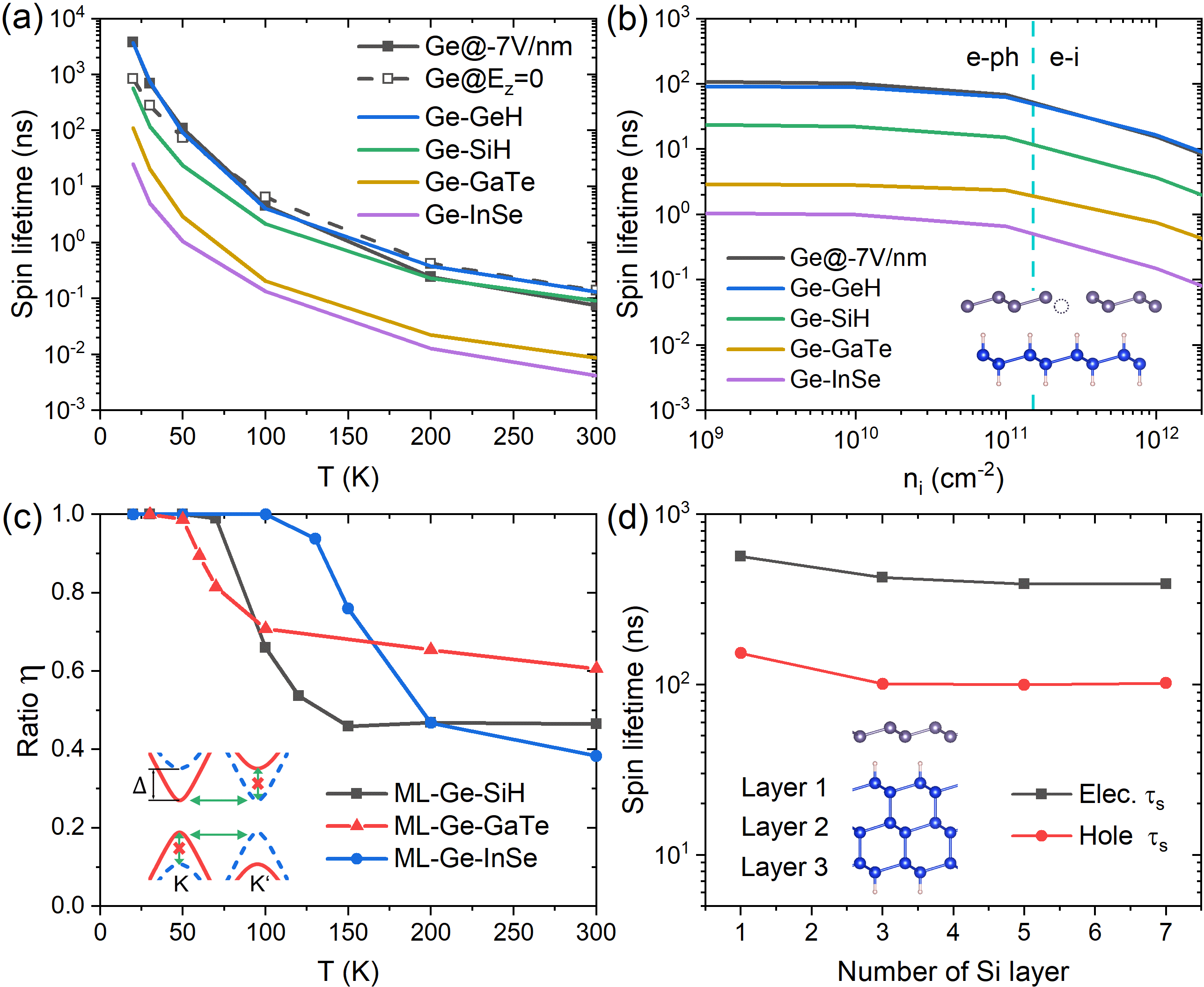}

\caption{The out-of-plane spin lifetime $\tau_{s}$ of intrinsic free-standing
and substrate-supported ML-Ge. (a) $\tau_{s}$ of ML-Ge under $E_{z}=0$,
-7 V/nm and substrate-supported ML-Ge as a function of temperature
without impurities. Here we show electron $\tau_{s}$ for intrinsic
ML-Ge systems except that hole $\tau_{s}$ is shown for ML-Ge-InSe,
since electron $\tau_{s}$ are longer than hole $\tau_{s}$ at low
$T$ except ML-Ge-InSe. (b) $\tau_{s}$ as a function of impurity
density $n_{i}$ at 50 K. The impurities are neutral ML-Ge vacancy
with 50$\%$ at higher positions and 50$\%$ at lower ones of a Ge
layer. The dashed vertical line corresponds to the impurity density
where e-ph and e-i scatterings contribute equally to spin relaxation
($n_{i,s}$). And e-ph (e-i) scattering is more dominant if $n_{i}~<~(>)~n_{i,s}$.
(c) The proportion of intervalley spin relaxation contribution $\eta$
of (electrons of) ML-Ge-SiH and (holes of) ML-Ge-InSe without impurities.
$\eta$ is defined as $\eta=\frac{\left(\tau_{s,z}^{\mathrm{inter}}\right)^{-1}}{\left(\tau_{s,z}^{\mathrm{inter}}\right)^{-1}+w\left(\tau_{s,z}^{\mathrm{intra}}\right)^{-1}}$,
where $\tau_{s,z}^{\mathrm{inter}}$ and $\tau_{s,z}^{\mathrm{intra}}$
are intervalley and intravalley spin lifetimes, corresponding to scattering
processes between $K$ and $K'$ valleys and within a single $K$
or $K'$ valley, respectively. $\eta$ being close to 1 or 0 corresponds
to dominant intervalley or intravalley spin relaxation, respectively.
$w$ is a weight factor related to what percentage
of total $S_{z}$ can be relaxed out by intravalley scattering itself.
$w$ being close to 0 and 1 correspond to the cases that intravalley scattering can
only relax a small part (0) and most of excess spin (1) respectively. In Supporting
Information Sec. SII, we give more details about definition of $w$.
(d) Electron and hole $\tau_{s}$ at 20 K of ML-Ge without impurities
on hydrogen-terminated multilayer Si, labeled as Si$_{n}$H with $n$
being number of Si layers. Si$_{n}$H is silicane if $n=1$, and hydrogen-terminated
Silicon (111) surface if $n=\infty$. \label{fig:T1}}
\end{figure*}

We then perform our first-principles density-matrix calculation~\cite{xu2020spin,xu2021ab,xu2021giant,xu2022spin} at proposed interfaces, and examine the role of electron-phonon coupling in spin relaxation of ML-Ge at different substrates.
Throughout this paper, we focus on out-of-plane $\tau_{s}$ of ML-Ge
systems, since their in-plane $\tau_{s}$ is too short and less interesting.
We compare out-of-plane $\tau_{s}$ due to e-ph scattering between
the free-standing ML-Ge (with/without an electric field) and ML-Ge
on different substrates in Fig.~\ref{fig:T1}a. Here we show electron
$\tau_{s}$ 
for most ML-Ge/substrate systems as intrinsic semiconductors, except hole
$\tau_{s}$ for the ML-Ge-InSe interface. This choice is because electron
$\tau_{s}$ are mostly longer than hole $\tau_{s}$ at low $T$ except
for the one at the ML-Ge-InSe interface; longer lifetime is often
more advantageous for spintronics applications. From Fig.~\ref{fig:T1}, we find that
$\tau_{s}$ of ML-Ge under $E_{z}=0$ and -7 V/nm are at the same order
of magnitude for a wide range of temperatures. The differences are
only considerable at low $T$, e.g, by 3-4 times at 20 K. On the other
hand, $\tau_{s}$ of supported ML-Ge are very sensitive to the specific
substrates. While $\tau_{s}$ of ML-Ge-GeH and ML-Ge-SiH have the
same order of magnitude as the free-standing ML-Ge, in particular
very close between ML-Ge-GeH and ML-Ge@-7 V/nm, $\tau_{s}$ of ML-Ge-GaTe
and ML-Ge-InSe are shorter by at least 1-2 orders of magnitude in
the whole temperature range. This separates the substrates into two
categories, i.e. with a weak effect (ML-Ge-GeH and ML-Ge-SiH) and
a strong effect (ML-Ge-GaTe and ML-Ge-InSe).

We further investigate the role of electron-impurity (e-i) scattering in spin relaxation under different substrates, by introducing defects in the material layer. 
We consider a common type of impurity - single neutral Ge
vacancy, whose formation energy was found relatively
low in previous theoretical studies\citep{padilha2016electronic,Ali2017}.
From Fig. \ref{fig:T1}b, we can see that $\tau_{s}$ of all five
systems decrease with impurity density $n_{i}$. Since carrier scattering
rates $\tau_{p}^{-1}$ (carrier lifetime $\tau_{p}$) increases (decrease)
with $n_{i}$, we then obtain $\tau_{s}$ decreases
with $\tau_{p}$'s decrease, an evidence of EY spin relaxation mechanism. Moreover,
we find that $\tau_{s}$ is sensitive to the type of the substrate with all values of $n_{i}$, and for each of four substrates, $\tau_{s}$ is reduced by
a similar amount with different $n_{i}$, from low density limit (10$^{9}$
cm$^{-2}$, where e-ph scattering dominates) to relatively high density
(10$^{12}$ cm$^{-2}$, where e-i scattering becomes more important).

Since the bands near the Fermi energy are composed of the Dirac cone
electrons around $K$ and $K'$ valleys in ML-Ge, spin relaxation
process arises from intervalley and intravalley e-ph scatterings.
We then examine relative intervalley spin relaxation
contribution $\eta$ (see its definition in the Fig.~\ref{fig:T1}
caption) in Fig.~\ref{fig:T1}c. $\eta$ being close to 1 or 0 corresponds to intervalley
or intravalley scattering being dominant in spin relaxation. $\eta$
becomes close to 1 below 70 K for electrons of ML-Ge-SiH, and below
120 K for holes of ML-Ge-InSe. This indicates that at low $T$ only
intervalley scattering processes are relevant to spin relaxation
in ML-Ge on substrates. This is a result of \textit{spin-valley locking} (SVL),
i.e. large SOC-induced band splittings lock up or down spin with a
particular K or K' valley~\cite{xu2021giant}. According to Fig.
\ref{fig:bands} and \ref{fig:T1}c, the SVL transition temperature
($T^{\mathrm{SVL}}$; below which the proportion of intervalley spin
relaxation rate $\eta$ is close to 1) seems approximately proportional
to SOC splitting energy $\Delta^{\mathrm{SOC}}$, e.g. for
electrons (CBM) of ML-Ge-GaTe and ML-Ge-SiH, and for holes (VBM) of ML-Ge-InSe,
$\Delta^{\mathrm{SOC}}$ are $\sim$15, $\sim$24 and 40 meV respectively,
while $T^{\mathrm{SVL}}$ are 50, 70 and 120 K respectively. 
As $\Delta^{\mathrm{SOC}}$
can be tuned by $E_{z}$ and the substrate, $T^{\mathrm{SVL}}$ can
be tuned simultaneously. Under SVL condition, spin or valley lifetime tends to be exceptionally long,
which is ideal for spin-/valley-tronic
applications.

Additionally, the studied substrates here are monolayer, while practically
multilayers or bulk are more common, thus it is necessary to understand
how $\tau_{s}$ changes with the number of substrate layers. In Fig.~\ref{fig:T1}d,
we show $\tau_{s}$ at 20 K of ML-Ge on hydrogen-terminated multilayer
Si, ML-Ge-Si$_{n}$H, with $n$ being number of Si layer. Si$_{n}$H
becomes hydrogen-terminated Silicon (111) surface if $n=\infty$.
We find that $\tau_{s}$ are changed by only 30$\%$-40$\%$ by increasing
$n$ from 1 to 3 and kept unchanged after $n\geq3$. For generality of our conclusion, we also test the layer dependence of a different substrate. We found the $\tau_{s}$ of ML-Ge on bilayer InSe ($n=2$) is changed by $\sim$8$\%$ compared to monolayer InSe at 20 K, even smaller change than the one at Si$_{n}$H substrates.
Given the disparate properties of these two substrates, we conclude using a monolayer is a reasonable choice 
for simulating the substrate effects on $\tau_{s}$ in this work.

\subsection*{The correlation of electronic structure and phonon properties to spin relaxation at different substrates}

\begin{figure*}[!ht]
\includegraphics[scale=0.21]{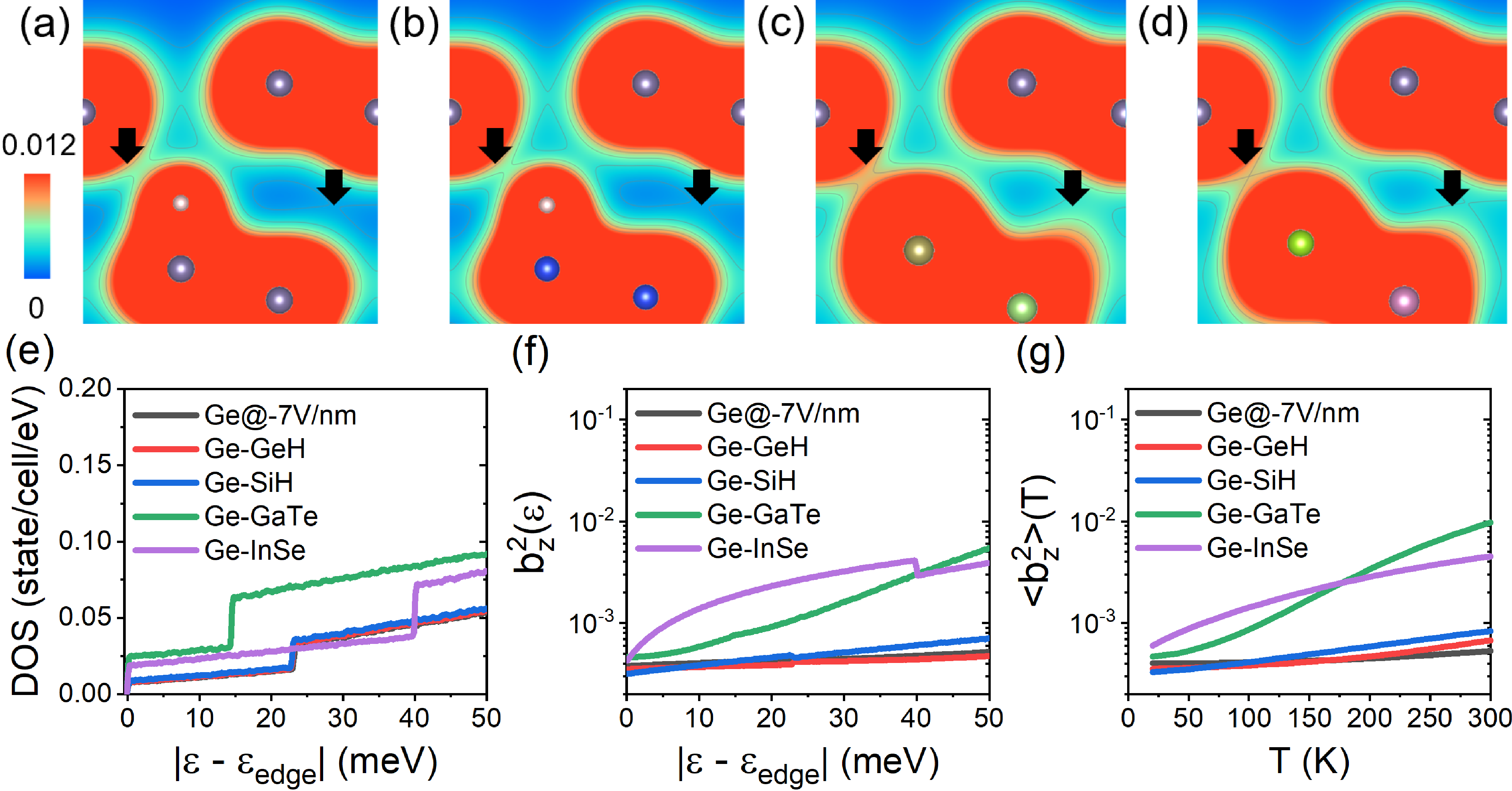}

\caption{Charge density, density of states (DOS), and spin mixing parameters
of free-standing and substrate-supported ML-Ge. Cross-section views
of charge density at interfaces of ML-Ge on (a) GeH, (b) SiH, (c)
GaTe, and (d) InSe. The Ge layers are above the substrate layers.
The unit of charge density is $e/\mathrm{bohr^{3}}$. Charge densities
in the regions pointed out by black arrows show significant differences
among different systems. (e) DOS and (f) energy-solved spin-mixing
parameter along $z$ axis $b_{z}^{2}\left(\varepsilon\right)$ of
ML-Ge under $E_{z}=$-7 V/nm and on different substrates. $\varepsilon_{\mathrm{edge}}$
is the band edge energy at the valence band maximum or conduction
band minimum. The step or sudden jump in the DOS curve corresponds
to the edge energy of the second conduction/valence band or the SOC-induced
splitting energy at $K$. (g) The temperature-dependent effective
spin-mixing parameter $\left\langle b_{z}^{2}\right\rangle $ of various
ML-Ge systems.\label{fig:electronic_quantities}}
\end{figure*}

We next analyze in detail the relevant physical quantities, and determine
the key factors responsible for substrate effects on spin relaxation.
We focus on results under low $T$ as spin relaxation properties are
superior at lower $T$ (the realization of SVL and longer $\tau_{s}$).

First, to have a qualitative understanding of the material-substrate
interaction strength, we show charge density distribution at the cross-section of interfaces in Fig.~\ref{fig:electronic_quantities}a-d.
It seems that four substrates can be categorized into two groups:
group A contains GeH and SiH with lower charge density distribution
in the bonding regions (pointed by the arrows); group B contains GaTe
and InSe with higher charge density distribution in the bonding regions.
In Fig. S5, we investigate the charge density change
$\Delta\rho^{e}$ (defined by the charge density difference between
interfaces and individual components). Consistent with Fig.~\ref{fig:electronic_quantities}, we find that $\Delta\rho^{e}$
for GaTe and InSe substrates overall has larger magnitude than the
one for GeH and SiH substrates. 
Therefore the material-substrate
interactions of group B seem stronger than those of group A. Intuitively,
we may expect that the stronger the interaction, the stronger the
substrate effect is. The FPDM simulations in Fig.~\ref{fig:T1}a-b
indeed show that the substrate effects of group B being stronger than
those of group A on $\tau_{s}$, consistent with the above intuition.

Next we examine electronic quantities closely related to spin-flip
scattering responsible to EY spin relaxation. Qualitatively, for a
state $k_{1}$, its spin-flip scattering rate $\tau_{s}^{-1}\left(k_{1}\right)$
is proportional to the number of its pair states $k_{2}$ allowing
spin-flip transitions between them. The number of pair states is approximately
proportional to density of states (DOS) around the energy of $k_{1}$.
Moreover, for EY mechanism, it is commonly assumed that spin relaxation
rate is proportional to the degree of mixture of spin-up and spin-down
states (along the $z$ direction here), so called ``spin-mixing''
parameter\cite{vzutic2004spintronics} $b_{z}^{2}$ (see its definition
in Sec. SII), i.e., $\tau_{s}^{-1}\propto\left\langle b_{z}^{2}\right\rangle $,
where $\left\langle b_{z}^{2}\right\rangle $ is the statistically
averaged spin mixing parameter as defined in Ref.~\citenum{xu2021giant}.
Therefore, we show DOS, energy-resolved spin-mixing $b_{z}^{2}\left(\varepsilon\right)$
and $\left\langle b_{z}^{2}\right\rangle $ as a function of temperature
in Fig.~\ref{fig:electronic_quantities}e-g.

We find that in Fig.~\ref{fig:electronic_quantities}e DOS of ML-Ge-GeH
and ML-Ge-SiH are quite close to that of ML-Ge@-7V/nm, while DOS of
ML-Ge-GaTe and ML-Ge-InSe are 50$\%$-100$\%$ higher around the band
edge. Such DOS differences are qualitatively explained
by the staggered potentials of ML-Ge-GaTe and ML-Ge-InSe being greater
than those of ML-Ge-GeH and ML-Ge-SiH according to the model Hamiltonian
proposed in Ref.~\citenum{kochan2017model}. In Fig.~\ref{fig:electronic_quantities}f-g,
$b_{z}^{2}$ of ML-Ge-GeH and ML-Ge-SiH are found similar to ML-Ge@-7
V/nm, and not sensitive to energy and temperature. On the contrast,
for ML-Ge-GaTe and ML-Ge-InSe, their $b_{z}^{2}\left(\varepsilon\right)$
and $\left\langle b_{z}^{2}\right\rangle $ increase rapidly with
energy and temperature. 
Specifically, we can see at 300 K, $\left\langle b_{z}^{2}\right\rangle $
of ML-Ge-GaTe and ML-Ge-InSe are about 4-20 times of the one of ML-Ge-GeH
and ML-Ge-SiH in Fig.~\ref{fig:electronic_quantities}g. Thus the
one order of magnitude difference of $\tau_{s}$ between group A (ML-Ge-GeH
and ML-Ge-SiH) and group B (ML-Ge-GaTe and ML-Ge-InSe) substrates
at 300 K can be largely explained by the substrate-induced changes
of DOS and $\left\langle b_{z}^{2}\right\rangle $. On the other hand,
at low $T$, e.g., at 50 K, $\left\langle b_{z}^{2}\right\rangle $
of ML-Ge-GaTe and ML-Ge-InSe are only about 1.5 and 2.5 times of the
ones of ML-Ge-GeH and ML-Ge-SiH, and DOS are only tens of percent
higher. However, there is still 1-2 order of magnitude difference
of $\tau_{s}$ between different substrates. Therefore, the substrate
effects on $\tau_{s}$ can not be fully explained by the changes of
$\left\langle b_{z}^{2}\right\rangle $ and DOS, in particular at
relatively low temperature.

\begin{figure}[!ht]
\sffamily
{\fontsize{13.5}{14.5}\selectfont {\small (}a{\small )}}
\resizebox{0.43\textwidth}{!}{
\begin{tabular}[t]{|c|c|c|}
\hline
Substrate & $\omega_{K}$ (meV) & Contribution\\
\hline
Ge@-7V/nm & 7.7 & 78$\%$\\
\hline
Ge-GeH & 6.9 & 70$\%$\\
\hline
Ge-SiH & 7.1 & 64$\%$\\
\hline
Ge-GaTe & 6.4 & 90$\%$\\
\hline
Ge-InSe & 7.2 & 99$\%$\\
\hline
\end{tabular}}

\textsf{\includegraphics[scale=0.48]{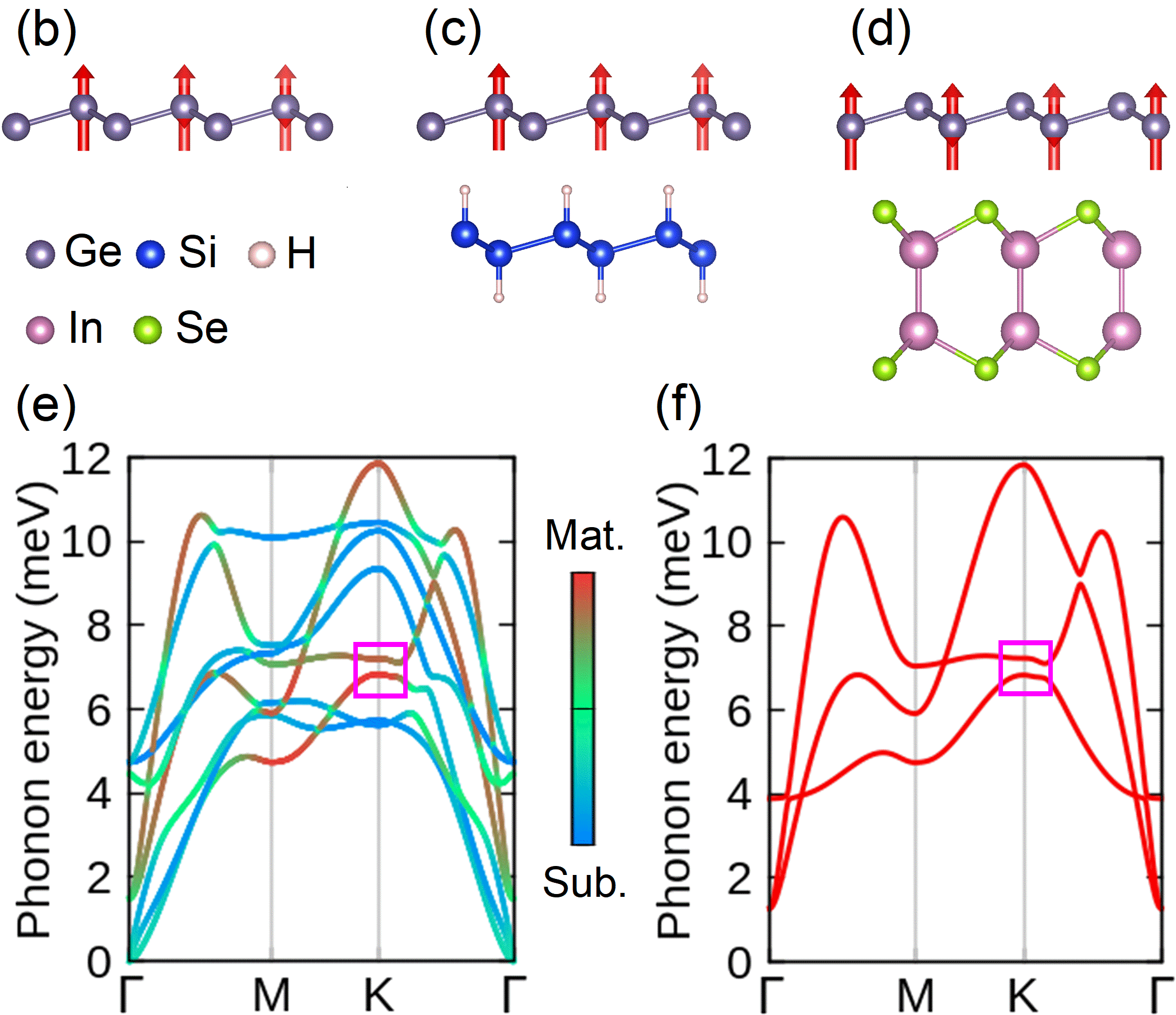}}

\textsf{\caption{(a) The phonon energy at wavevector $K$ of the mode that contributes
the most to spin relaxation, and the percentage of its contribution
for various systems at 20 K. We consider momentum transfer $K$, as spin relaxation
is fully determined by intervalley processes between $K$ and $K'$
valleys. (b), (c) and (d) Typical vibrations of atoms in 3$\times$3
supercells of (b) ML-Ge@-7 V/nm, (c) ML-Ge-SiH, and (d) ML-Ge-InSe
of the most important phonon mode at $K$ around 7 meV (shown in (a)). The red arrows represent
displacement. The atomic displacements smaller than 10$\%$ of the
strongest are not shown. (e) The layer-projected
phonon dispersion of ML-Ge-InSe within 12 meV. The red and blue colors
correspond to the phonon displacements mostly contributed from
the material (red) and substrate layer (blue) respectively. The green color means
the contribution to the phonon
displacements from the material and substrate layers are similar. The purple boxes highlight the two most
important phonon modes around $K$ for spin relaxation.(f) Phonon
dispersion of ML-Ge-InSe within 12 meV with substrate atoms (InSe)
being fixed at equilibrium structure and only Ge atoms are allowed
to vibrate.\label{fig:phonons}}}
\end{figure}

We then examine if substrate-induced
modifications of phonon can explain the changes of spin relaxation at different substrates, especially at low $T$. 
We emphasize that at low $T$, since spin
relaxation is fully determined by intervalley processes (Fig.~\ref{fig:T1}c),
the related phonons 
are mostly
close to wavevector $K$. From Fig.~\ref{fig:phonons}, we find that the most
important phonon mode for spin relaxation at low
$T$ has several similar features: (i) 
It contributes to more than 60$\%$ of spin relaxation 
(see Fig \ref{fig:phonons}a).
(ii) Its energy is around 7 meV in the table of Fig.~\ref{fig:phonons}a.
(iii) Its vibration is flexural-like, i.e., atoms mostly vibrate along the out-of-plane
direction as shown in Fig.~\ref{fig:phonons}b-d. Moreover, for this mode, the substrate atoms have negligible thermal vibration amplitude  compared to the one of the materials atoms. 
This is also confirmed in the layer-projected phonon dispersion of ML-Ge-InSe
in Fig.~\ref{fig:phonons}e. The purple box highlights the critical phonon mode around $K$, with most contribution from the material layer.
(iv) The critical phonon
mode does not couple with the substrate strongly, since its vibration
frequency does not change much when substrate atoms are fixed (by
comparing Fig.~\ref{fig:phonons}e with f). 
We thus conclude that the substrate-induced modifications
of phonons and thermal vibrations of substrate atoms seem not important
for spin relaxation at low $T$ (e.g. below 20 K).

\begin{figure*}[!ht]
\includegraphics[scale=0.7]{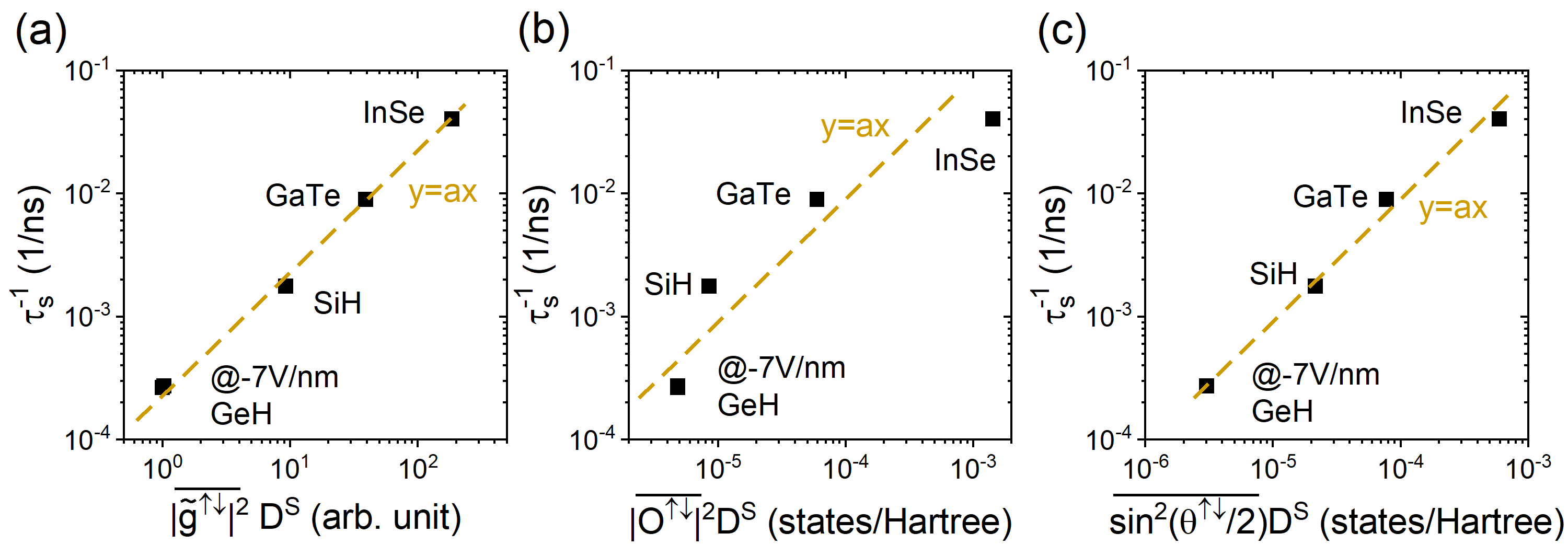}

\caption{The relation between $\tau_{s}^{-1}$ and the averaged
modulus square of spin-flip e-ph matrix elements $\overline{|\widetilde{g}^{\uparrow\downarrow}|^{2}}$,
of spin-flip overlap matrix elements $\overline{|o^{\uparrow\downarrow}|^{2}}$
and $\overline{\mathrm{sin}^{2}\left(\theta^{\uparrow\downarrow}/2\right)}$
multiplied by the scattering density of states $D^{S}$ at 20 K. See
the definition of $\overline{|\widetilde{g}^{\uparrow\downarrow}|^{2}}$,
$\overline{|o^{\uparrow\downarrow}|^{2}}$ and $D^{S}$ in Eq. \ref{eq:gsf2},
\ref{eq:osf2} and \ref{eq:DS} respectively. $\theta^{\uparrow\downarrow}$
is the spin-flip angle between two electronic states. For two states
$\left(k,n\right)$ and $\left(k',n'\right)$ with opposite spin directions,
$\theta^{\uparrow\downarrow}$ is the angle between $-{\bf S}_{kn}^{\mathrm{exp}}$
and ${\bf S}_{k'n'}^{\mathrm{exp}}$. $\overline{\mathrm{sin}^{2}\left(\theta^{\uparrow\downarrow}/2\right)}$
is defined in Eq. \ref{eq:sin2}. The variation of $D^{S}$ among
different substrates is at most three times, much weaker than the
variations of $\tau_{s}^{-1}$ and other quantities shown here.\label{fig:spin-flip_matrix_elements}}
\end{figure*}

Therefore, neither the simple electronic quantities $\left\langle b^{2}\right\rangle $
and DOS nor the phonon properties can explain the substrate effects
on spin relaxation at low $T$. 


\subsection*{The determining factors of spin relaxation derived from spin-flip matrix elements}

On the other hand, with a simplified picture of spin-flip transition
by the Fermi's Golden Rule, the scattering rate is proportional to
the modulus square of the scattering matrix elements. For a further mechanistic understanding, we
turn to examine the modulus square of the spin-flip matrix elements,
and compare their qualitative trend with our FPDM simulations.
Note that most matrix elements are irrelevant to spin relaxation and
we need to pick the ``more relevant'' ones, by defining a statistically-averaged function.
Therefore, we propose an effective band-edge-averaged spin-flip matrix element 
$\overline{|\widetilde{g}^{\uparrow\downarrow}|^{2}}$
(Eq.~\ref{eq:gsf2}). Here the spin-flip matrix element can be for general scattering processes; in the following we focus on e-ph process for simplicity.  We also propose a so-called scattering density
of states $D^{S}$ in Eq. \ref{eq:DS}, which measures the density
of spin-flip transitions and can be roughly regarded as a weighted-averaged value of the usual DOS. Based on the generalized Fermi's golden rule, we
approximately have $\tau_{s}^{-1}\propto\overline{|\widetilde{g}^{\uparrow\downarrow}|^{2}}D^{S}$
for EY spin relaxation (see the discussions above Eq. \ref{eq:taus_approx}
in ``Methods" section).

As shown in Fig.~\ref{fig:spin-flip_matrix_elements}a, $\tau_{s}^{-1}$
is almost linearly proportional to $\overline{|\widetilde{g}^{\uparrow\downarrow}|^{2}}D^{S}$
at 20 K. 
As the variation of $D^{S}$ among ML-Ge
on different substrates is at most three times (see Fig.~\ref{fig:electronic_quantities}e
and Fig.~S6), which is much weaker than the large variation of $\tau_{s}^{-1}$,
this indicates that the substrate-induced change of $\tau_{s}$ is
mostly due to the substrate-induced change of spin-flip matrix elements.
Although $\overline{|\widetilde{g}^{\uparrow\downarrow}|^{2}}$ was
often considered approximately proportional to $\left\langle b^{2}\right\rangle $,
resulting in $\tau_{s}^{-1}\propto\left\langle b^{2}\right\rangle $,
our results in Fig.~\ref{fig:electronic_quantities} in the earlier section indicate that such simple approximation is not applicable here, especially inadequate of explaining
substrate dependence of $\tau_{s}$ at low $T$. 

To find out the reason why $\overline{|\widetilde{g}^{\uparrow\downarrow}|^{2}}$
for different substrates are so different, 
we first examine the averaged spin-flip wavefunction overlap 
$\overline{|o^{\uparrow\downarrow}|^{2}}$ (with the reciprocal lattice vector ${\bf G}=0$),  closely related to $\overline{|\widetilde{g}^{\uparrow\downarrow}|^{2}}$ (Eq.~\ref{eq:oG} and Eq.~\ref{eq:Elliot}).
From Fig.~\ref{fig:spin-flip_matrix_elements}b, $\tau_{s}^{-1}$
and $\overline{|o^{\uparrow\downarrow}|^{2}}$ have the same trend,
which implies $\overline{|\widetilde{g}^{\uparrow\downarrow}|^{2}}$
and $\overline{|o^{\uparrow\downarrow}|^{2}}$ may have the same trend. However, in general, the $\bf{G}\neq 0$ elements of $\overline{|o^{\uparrow\downarrow}|^{2}}$ may be important as well, which can not be unambiguously evaluated here. (See detailed
discussions in the subsection ``Spin-flip e-ph and overlap matrix
element'' in the ``Methods'' section). 

To have deeper intuitive understanding, we then propose an important electronic quantity
for intervalley spin-flip scattering - the spin-flip angle $\theta^{\uparrow\downarrow}$
between two electronic states. For two states $\left(k_{1},n_{1}\right)$
and $\left(k_{2},n_{2}\right)$ with opposite spin directions, $\theta^{\uparrow\downarrow}$
is the angle between $-{\bf S}_{k_{1}n_{1}}^{\mathrm{exp}}$ and ${\bf S}_{k_{2}n_{2}}^{\mathrm{exp}}$ or equivalently the angle between $-{\bf B}_{k_{1}}^{\mathrm{in}}$ and ${\bf B}_{k_{2}}^{\mathrm{in}}$.

The motivation of examining $\theta^{\uparrow\downarrow}$ is that: 
Suppose two wavevectors ${\bf k}_{1}$ and ${\bf k}_{2}={\bf -k}_{1}$ are in two opposite valleys $Q$ and -$Q$ respectively and there is a pair of bands, which are originally Kramers degenerate but splitted by ${\bf B}^{\mathrm{in}}$.
Due to time-reversal symmetry, we have ${\bf B}^{\mathrm{in}}_{k_1}=-{\bf B}^{\mathrm{in}}_{k_2}$, which means the two states at the same band $n$ at ${\bf k_1}$ and ${\bf k_2}$ have opposite spins and $\theta^{\uparrow\downarrow}$ between them is zero.
Therefore, the matrix element of operator $\widehat{A}$ between states $\left(k_{1},n\right)$ and $\left(k_{2},n\right)$ - $A_{k_{1}n,k_{2}n}$ is a spin-flip one and we name it as $A^{\uparrow\downarrow}_{k_{1}k_{2}}$.
According to Ref. \citenum{yafet1963g}, with time-reversal symmetry, $A^{\uparrow\downarrow}_{k_{1}k_{2}}$ is exactly zero.
In general, for another wavevector ${\bf k_3}$ within valley -$Q$ but not ${\bf -k}_{1}$, $A^{\uparrow\downarrow}_{k_{1}k_{3}}$ is usually non-zero. 
One critical quantity that determines the intervalley spin-flip matrix element $A^{\uparrow\downarrow}_{k_{1}k_{3}}$ for a band within the pair introduced above is $\theta^{\uparrow\downarrow}_{k_1k_3}$.
Based on time-independent perturbation theory, we can prove that
$\left|A^{\uparrow\downarrow}\right|$ between two states is approximately proportional to $\left|\sin\left(\theta^{\uparrow\downarrow}/2\right)\right|$.
The derivation is given in subsection ``Spin-flip angle $\theta^{\uparrow\downarrow}$ for intervalley spin relaxation" in ``Methods" section.

As shown in Fig.~\ref{fig:spin-flip_matrix_elements}c,
$\tau_{s}^{-1}$ of ML-Ge on different substrates at 20 K is almost linearly proportional to $\overline{\sin^{2}\left(\theta^{\uparrow\downarrow}/2\right)}D^{S}$,
where $\overline{\sin^{2}\left(\theta^{\uparrow\downarrow}/2\right)}$
is the statistically-averaged modulus square of $\sin\left(\theta^{\uparrow\downarrow}/2\right)$.
This indicates that the relation $\overline{|\widetilde{g}^{\uparrow\downarrow}|^{2}}\propto\overline{\sin^{2}\left(\theta^{\uparrow\downarrow}/2\right)}$
is nearly perfectly satisfied at low $T$, where intervalley processes dominate
spin relaxation. We additionally show the relations between $\tau_{s}^{-1}$
and $\overline{|\widetilde{g}^{\uparrow\downarrow}|^{2}}D^{S}$, $\overline{|o^{\uparrow\downarrow}|^{2}}D^{S}$
and $\overline{\sin^{2}\left(\theta^{\uparrow\downarrow}/2\right)}D^{S}$ at 300 K 
in Fig. S7. Here the trend of $\tau_{s}^{-1}$ is still approximately
captured by the trends of $\overline{|\widetilde{g}^{\uparrow\downarrow}|^{2}}D^{S}$,
$\overline{|o^{\uparrow\downarrow}|^{2}}D^{S}$ and $\overline{\sin^{2}\left(\theta^{\uparrow\downarrow}/2\right)}D^{S}$, although not perfectly linear as at low $T$.

Since $\theta^{\uparrow\downarrow}$ is defined
by ${\bf S}^{\mathrm{exp}}$ at different states, $\tau_{s}$ is highly correlated with ${\bf S}^{\mathrm{exp}}$ and more specifically with the anisotropy of ${\bf S}^{\mathrm{exp}}$ (equivalent to the anisotropy of ${\bf B}^{\mathrm{in}}$). 
Qualitatively, the larger anisotropy of ${\bf S}^{\mathrm{exp}}$ leads to smaller $\theta^{\uparrow\downarrow}$ and longer $\tau_{s}$ along the high-spin-polarization direction.
This finding may be applicable
to spin relaxation in other materials whenever intervalley spin-flip
scattering dominates or spin-valley locking exists, e.g., in TMDs\citep{dey2017gate},
Stanene\citep{tao2019two}, 2D hybrid perovskites with persistent spin helix\citep{zhang2022room}, etc.

At the end, we briefly discuss the substrate effects
on in-plane spin relaxation ($\tau_{s,x}$), whereas only out-of-plane spin relaxation was discussed earlier. From Table SI, we find
that $\tau_{s,x}$ of ML-Ge@-7V/nm and supported ML-Ge are significantly
(e.g., two orders of magnitude) shorter than free-standing ML-Ge,
but the differences between $\tau_{s,x}$ of ML-Ge on different substrates
are relatively small (within 50$\%$). This is because: With a non-zero $E_{z}$ or a substrate, the inversion
symmetry broken induces strong out-of-plane internal magnetic field
$B_{z}^{\mathrm{in}}$ (\textgreater 100 Tesla), so that the excited
in-plane spins will precess rapidly about $B_{z}^{\mathrm{in}}$.
The spin precession significantly affects spin decay and the main
spin decay mechanism becomes DP or free induction decay mechanism\cite{wu2010spin}
instead of EY mechanism. For both DP and free induction decay mechanisms\cite{wu2010spin,xu2022spin}, $\tau_{s,x}$ decreases with the fluctuation amplitude (among different k-points) of the ${\bf B}^{\mathrm{in}}$ components perpendicular to the $x$ direction.
As the fluctuation amplitude of $B_{z}^{\mathrm{in}}$ of ML-Ge@-7V/nm and supported ML-Ge is large (Table SI; much greater than the one of $B_{y}^{\mathrm{in}}$), their $\tau_{s,x}$ can be much shorter than the value of ML-Ge at zero electric field when EY mechanism dominates. Moreover,
since the fluctuation amplitude of $B_{z}^{\mathrm{in}}$ of ML-Ge on different substrates has
the same order of magnitude (Table SI), $\tau_{s,x}$ of ML-Ge on different substrates are similar.

\section*{Conclusions}

In this paper, we systematically investigate how spin relaxation of
strong SOC Dirac materials is affected by different insulating substrates,
using germanene as a prototypical example. Through FPDM simulations
of $\tau_{s}$ of free-standing and substrate supported ML-Ge, we
show that substrate effects on $\tau_{s}$ can differ orders of magnitude
among different substrates. Specifically, $\tau_{s}$ of ML-Ge-GeH
and ML-Ge-SiH have the same order of magnitude as free-standing ML-Ge,
but $\tau_{s}$ of ML-Ge-GaTe and ML-Ge-InSe are significantly shortened
by 1-2 orders with temperature increasing from 20 K to 300 K.

Although simple electronic quantities including charge densities,
DOS and spin mixing $\left\langle b_{z}^{2}\right\rangle $ qualitatively
explain the much shorter lifetime of ML-Ge-GaTe/InSe compared to ML-Ge-GeH/SiH
in the relatively high $T$ range, we find they cannot
explain the large variations of $\tau_{s}$ among substrates at low
$T$ (i.e. tens of K). We point out that spin relaxation in ML-Ge
and its interfaces at low $T$ is dominated by intervalley scattering
processes. However, the substrate-induced modifications of phonons and
thermal vibrations of substrates seem to be not important.
Instead, the substrate-induced changes of the anisotropy
of ${\bf S}^{\mathrm{exp}}$ or the spin-flip angles $\theta^{\uparrow\downarrow}$
which changes the spin-flip matrix elements, are much more crucial.
$\theta^{\uparrow\downarrow}$ is at the first time
proposed in this article to the best of our knowledge, and is found to be
a useful electronic quantity for predicting trends of spin relaxation when intervalley
spin-flip scattering dominates.

Our theoretical study showcases the systematic investigations of the
critical factors determining the spin relaxation in 2D Dirac materials.
More importantly we pointed out the sharp distinction of substrate
effects on strong SOC materials to the effects on weak SOC ones, providing
valuable insights and guidelines for optimizing spin relaxation in
materials synthesis and control.

\section*{Methods}

\subsection*{First-Principles Density-Matrix Dynamics for Spin Relaxation}

We solve the quantum master equation of density matrix
$\rho\left(t\right)$ as the following:\citep{xu2021ab} 
\begin{align}
\frac{d\rho_{12}\left(t\right)}{dt}= & \left[H_{e},\rho\left(t\right)\right]_{12}+\nonumber \\
 & \left(\begin{array}{c}
\frac{1}{2}\sum_{345}\left\{ \begin{array}{c}
\left[I-\rho\left(t\right)\right]_{13}P_{32,45}\rho_{45}\left(t\right)\\
-\left[I-\rho\left(t\right)\right]_{45}P_{45,13}^{*}\rho_{32}\left(t\right)
\end{array}\right\} \\
+H.C.
\end{array}\right),\label{eq:master}
\end{align}
Eq.~\ref{eq:master} is expressed in the Schrödinger picture, where
the first and second terms on the right side of the equation relate
to the coherent dynamics, which can lead to Larmor precession, and
scattering processes respectively. The first term is unimportant for
out-of-plane spin relaxation in ML-Ge systems, since Larmor precession
is highly suppressed for the excited spins along the out-of-plane
or $z$ direction due to high spin polarization along $z$ direction.
The scattering processes induce spin relaxation via the SOC. $H_{e}$
is the electronic Hamiltonian. $\left[H,\rho\right]\,=\,H\rho-\rho H$.
H.C. is Hermitian conjugate. The subindex, e.g., ``1'' is the combined
index of $\textbf{k}$-point and band. $P=P^{\mathrm{e-ph}}+P^{\mathrm{e-i}}$
is the generalized scattering-rate matrix considering e-ph and e-i
scattering processes.

For the e-ph scattering\citep{xu2021ab},
\begin{align}
P_{1234}^{\mathrm{e-ph}}= & \sum_{q\lambda\pm}A_{13}^{q\lambda\pm}A_{24}^{q\lambda\pm,*},\\
A_{13}^{q\lambda\pm}= & \sqrt{\frac{2\pi}{\hbar}}g_{12}^{q\lambda\pm}\sqrt{\delta_{\sigma}^{G}\left(\epsilon_{1}-\epsilon_{2}\pm\omega_{q\lambda}\right)}\sqrt{n_{q\lambda}^{\pm}},
\end{align}
where $q$ and $\lambda$ are phonon wavevector and mode, $g^{q\lambda\pm}$
is the e-ph matrix element, resulting from the absorption ($-$) or
emission ($+$) of a phonon, computed with self-consistent SOC from
first-principles,\citep{giustino2017electron} $n_{q\lambda}^{\pm}=n_{q\lambda}+0.5\pm0.5$
in terms of phonon Bose factors $n_{q\lambda}$, and $\delta_{\sigma}^{G}$
represents an energy conserving $\delta$-function broadened to a
Gaussian of width $\sigma$.

For electron-impurity scattering\citep{xu2021ab},
\begin{align}
P_{1234}^{\mathrm{e-i}}= & A_{13}^{i}A_{24}^{i,*},\\
A_{13}^{i}= & \sqrt{\frac{2\pi}{\hbar}}g_{13}^{i}\sqrt{\delta_{\sigma}^{G}\left(\epsilon_{1}-\epsilon_{3}\right)}\sqrt{n_{i}V_{\mathrm{cell}}},
\end{align}
where $n_{i}$ and $V_{\mathrm{cell}}$ are impurity density and unit
cell volume, respectively. $g^{i}$ is the e-i matrix element computed by the supercell method and
is discussed in the next subsection.

Starting from an initial density matrix $\rho\left(t_{0}\right)$
prepared with a net spin, we evolve $\rho\left(t\right)$ through
Eq. \ref{eq:master} for a long enough time, typically from hundreds
of ps to a few $\mu$s. We then obtain spin observable $S\left(t\right)$
from $\rho\left(t\right)$ (Eq. S1) and extract spin lifetime $\tau_{s}$
from $S\left(t\right)$ using Eq. S2.

\subsection*{Computational details}

The ground-state electronic structure, phonons, as
well as electron-phonon and electron-impurity (e-i) matrix elements
are firstly calculated using density functional theory (DFT) with
relatively coarse $k$ and $q$ meshes in the DFT plane-wave code
JDFTx\citep{sundararaman2017jdftx}. Since all substrates have hexagonal
structures and their lattice constants are close to germanene's, the
heterostructures are built simply from unit cells of two systems.
The lattice mismatch values are within 1$\%$ for GeH, GaTe and InSe
substrates but about 3.5$\%$ for the SiH substrate. All heterostructures
use the lattice constant 4.025 $\mathrm{\AA}$ of free-standing ML-Ge
relaxed with Perdew-Burke-Ernzerhof exchange-correlation functional\citep{perdew1996generalized}.
The internal geometries are fully relaxed using the DFT+D3 method
for van der Waals dispersion corrections\citep{grimme2010consistent}.
We use Optimized Norm-Conserving Vanderbilt (ONCV) pseudopotentials\citep{hamann2013optimized}
with self-consistent spin-orbit coupling throughout, which we find
converged at a kinetic energy cutoff of 44, 64, 64, 72 and 66 Ry for
free-standing ML-Ge, ML-Ge-GeH, ML-Ge-SiH, ML-Ge-GaTe and ML-Ge-InSe
respectively. The DFT calculations use 24$\times$24 $k$ meshes.
The phonon calculations employ $3\times3$ supercells through finite
difference calculations. We have checked the supercell size convergence
and found that using $6\times6$ supercells lead to very similar results
of phonon dispersions and spin lifetimes. For all systems, the Coulomb
truncation technique\citep{ismail2006truncation} is employed to accelerate
convergence with vacuum sizes. The vacuum sizes are 20 bohr (additional
to the thickness of the heterostructures) for all heterostructures
and are found large enough to converge the final results of spin lifetimes.
The electric field along the non-periodic direction is applied as a ramp potential.

For the e-i scattering, we assume impurity density
is sufficiently low and the average distance between neighboring impurities
is sufficiently long so that the interactions between impurities are
negligible, i.e. at the dilute limit.  The e-i matrix $g^{i}$ between state $\left(k,n\right)$
and $\left(k',n'\right)$ is $g_{kn,k'n'}^{i}=\left\langle kn\right|V^{i}-V^{0}\left|k'n'\right\rangle $,
where $V^{i}$ is the potential of the impurity system and $V^{0}$
is the potential of the pristine system. $V^{i}$ is computed with
SOC using a large supercell including a neutral impurity that simulates the dilute limit where impurity and its periodic replica do not interact. To speed
up the supercell convergence, we used the potential alignment method
developed in Ref. \citenum{sundararaman2017first}. We use 5$\times$5
supercells, which have shown reasonable convergence (a few percent
error of the spin lifetime).

We then transform all quantities from plane wave
basis to maximally localized Wannier function basis\citep{marzari1997maximally},
and interpolate them\citep{PhononAssisted,giustino2017electron,GraphiteHotCarriers,brown2017experimental,NitrideCarriers,TAparameters}
to substantially finer k and q meshes. The fine $k$ and $q$ meshes
are $384\times384$ and $576\times576$ for simulations at 300 K and
100 K respectively and are finer at lower temperature, e.g., $1440\times1440$
and $2400\times2400$ for simulations at 50 K and 20 K respectively.

The real-time dynamics simulations are done with
our own developed DMD code interfaced to JDFTx. The energy-conservation
smearing parameter $\sigma$ is chosen to be comparable or smaller
than $k_{B}T$ for each calculation, e.g., 10 meV, 5 meV, 3.3 meV
and 1.3 meV at 300 K, 100 K, 50 K and 20 K respectively.

\subsection*{Analysis of Elliot-Yafet spin lifetime}

In order to analyze the results from real-time first-principles
density-matrix dynamics (FPDM), we compare them with simplified mechanistic
models as discussed below. According to Ref.~\cite{xu2020spin},
if a solid-state system is close to equilibrium (but not at equilibrium)
and its spin relaxation is dominated by EY mechanism, its spin lifetime
$\tau_{s}$ due to the e-ph scattering satisfies (for simplicity the
band indices are dropped)

\begin{align}
\tau_{s}^{-1}\propto & \frac{N_{k}^{-2}}{\chi}\sum_{kq\lambda}\left\{ \begin{array}{c}
|g_{k,k-q}^{\uparrow\downarrow,q\lambda}|^{2}n_{q\lambda}f_{k-q}\left(1-f_{k}\right)\\
\delta\left(\epsilon_{k}-\epsilon_{k-q}-\omega_{q\lambda}\right)
\end{array}\right\} ,\label{eq:FGR-1}\\
\chi= & N_{k}^{-1}\sum_{k}f_{k}\left(1-f_{k}\right),\label{eq:FGR-2}
\end{align}

where $f$ is Fermi-Dirac function. $\omega_{q\lambda}$
and $n_{q\lambda}$ are phonon energy and occupation of phonon mode
$\lambda$ at wavevector $q$. $g^{\uparrow\downarrow}$ is the spin-flip
e-ph matrix element between two electronic states of opposite spins.
We will further discuss $g^{\uparrow\downarrow}$ in the next subsection.

According to Eq. \ref{eq:FGR-1} and \ref{eq:FGR-2},
$\tau_{s}^{-1}$ is proportional to $|g_{q}^{\uparrow\downarrow}|^{2}$
and also the density of the spin-flip transitions. Therefore we propose
a temperature ($T$) and chemical potential ($\mu_{F,c}$) dependent
effective modulus square of the spin-flip e-ph matrix element $\overline{|\widetilde{g}^{\uparrow\downarrow}|^{2}}$
and a scattering density of states $D^{\mathrm{S}}$ as 
\begin{align}
\overline{|\widetilde{g}^{\uparrow\downarrow}|^{2}}= & \frac{\sum_{kq}\mathrm{w}_{k,k-q}\sum_{\lambda}|g_{k,k-q}^{\uparrow\downarrow,q\lambda}|^{2}n_{q\lambda}}{\sum_{kq}\mathrm{w}_{k,k-q}},\label{eq:gsf2}\\
D^{\mathrm{S}}= & \frac{N_{k}^{-2}\sum_{kq}\mathrm{w}_{k,k-q}}{N_{k}^{-1}\sum_{k}f_{k}\left(1-f_{k}\right)},\label{eq:DS}\\
\mathrm{w}_{k,k-q}= & f_{k-q}\left(1-f_{k}\right)\delta\left(\epsilon_{k}-\epsilon_{k-q}-\omega_{c}\right),
\end{align}

where $\omega_{c}$ is the characteristic phonon
energy specified below, and $\mathrm{w}_{k,k-q}$ is the weight function.
The matrix element modulus square is weighted by $n_{q\lambda}$ according
to Eq. \ref{eq:FGR-1} and \ref{eq:FGR-2}. This rules out high-frequency
phonons at low $T$ which are not excited. $\omega_{c}$ is chosen
as 7 meV at 20 K based on our analysis of phonon-mode-resolved contribution
to spin relaxation. $\mathrm{w}_{k,k-q}$ selects transitions between
states separated by $\omega_{c}$ and around the band edge or $\mu_{F,c}$,
which are ``more relevant'' transitions to spin relaxation.

$D^{\mathrm{S}}$ can be regarded as an effective
density of spin-flip e-ph transitions satisfying energy conservation
between one state and its pairs. When $\omega_{c}=0$, we have $D^{\mathrm{S}}=\int d\epsilon\left(-\frac{df}{d\epsilon}\right)D^{2}\left(\epsilon\right)/\int d\epsilon\left(-\frac{df}{d\epsilon}\right)D\left(\epsilon\right)$
with $D\left(\epsilon\right)$ density of electronic states (DOS).
So $D^{\mathrm{S}}$ can be roughly regarded as a weighted-averaged
DOS with weight $\left(-\frac{df}{d\epsilon}\right)D\left(\epsilon\right)$.

With $\overline{|\widetilde{g}^{\uparrow\downarrow}|^{2}}$
and $D^{\mathrm{S}}$, we have the approximate relation for spin relaxation
rate, 
\begin{align}
\tau_{s}^{-1}\propto & \overline{|\widetilde{g}^{\uparrow\downarrow}|^{2}}D^{\mathrm{S}}.\label{eq:taus_approx}
\end{align}

\subsection*{Spin-flip e-ph and overlap matrix element}

In the mechanistic model of Eq.~\ref{eq:FGR-1}
in the last section, the spin-flip e-ph matrix element between two
electronic states of opposite spins at wavevectors ${\bf k}$ and
${\bf k-q}$ of phonon mode $\lambda$ reads\cite{giustino2017electron}

\begin{align}
g_{kk-q}^{\uparrow\downarrow,q\lambda}= & \left\langle u_{k}^{\uparrow(\downarrow)}\right|\Delta_{q\lambda}v^{\mathrm{KS}}\left|u_{k-q}^{\downarrow\left(\uparrow\right)}\right\rangle ,\label{eq:gsf}\\
\Delta_{q\lambda}v^{\mathrm{KS}}= & \sqrt{\frac{\hbar}{2\omega_{q\lambda}}}\sum_{\kappa\alpha}\frac{e_{\kappa\alpha,q\lambda}\partial_{\kappa\alpha q}v^{\mathrm{KS}}}{\sqrt{m_{\kappa}}},\\
\partial_{\kappa\alpha q}v^{\mathrm{KS}}= & \sum_{l}e^{i{\bf q}\cdot{\bf R}_{l}}\frac{\partial V^{\mathrm{KS}}}{\partial\tau_{\kappa\alpha}}|_{{\bf r}-{\bf R}_{l}},\\
V^{\mathrm{KS}}= & V+\frac{\text{\ensuremath{\hbar}}}{4m^{2}c^{2}}\nabla_{{\bf r}}V\times{\bf p}\cdot\sigma,\label{eq:VKS}
\end{align}

where $u_{k}^{\uparrow(\downarrow)}$ is the periodic
part of the Bloch wavefunction of a spin-up (spin-down) state at wavevector
${\bf k}$. $\kappa$ is the index of ion in the unit cell. $\alpha$
is the index of a direction. ${\bf R}_{l}$ is a lattice vector. $V$
is the spin-independent part of the potential. ${\bf p}$ is the momentum
operator. $\sigma$ is the Pauli operator.

From Eqs. \ref{eq:gsf}-\ref{eq:VKS}, $g^{\uparrow\downarrow}$
can be separated into two parts,

\begin{align}
g^{\uparrow\downarrow}= & g^{\mathrm{E}}+g^{\mathrm{Y}},\label{eq:E+Y}
\end{align}

where $g^{\mathrm{E}}$ and $g^{\mathrm{Y}}$ correspond
to the spin-independent and spin-dependent parts of $V^{\mathrm{KS}}$
respectively, called Elliot and Yafet terms of the spin-flip scattering
matrix elements respectively.\cite{wu2010spin}

Generally speaking, both the Elliot and Yafet terms
are important; for the current systems $\tau_{s}$ with and without
Yafet term have the same order of magnitude.
For example, $\tau_{s}$ of ML-Ge-GeH and ML-Ge-SiH without the Yafet
term are about 100$\%$ and 70$\%$ of $\tau_{s}$ with the Yafet
term at 20 K.
Therefore, for qualitative discussion of $\tau_{s}$ of ML-Ge
on different substrates (the quantitative calculations of $\tau_{s}$ are performed by FPDM introduced earlier), it is reasonable to focus on the Elliot term $g^{\mathrm{E}}$
and avoid the more complicated Yafet term $g^{\mathrm{Y}}$.

Define $V_{q\lambda}^{\mathrm{E}}$ as the spin-independent
part of $\Delta_{q\lambda}v^{\mathrm{KS}}$, so that $g^{\mathrm{E}}=\left\langle u_{k}^{\uparrow(\downarrow)}\right|V_{q\lambda}^{\mathrm{E}}\left|u_{k-q}^{\downarrow\left(\uparrow\right)}\right\rangle $.
Expanding $V_{q\lambda}^{\mathrm{E}}$ as $\sum_{G}\widetilde{V}_{q\lambda}^{\mathrm{E}}\left({\bf G}\right)e^{i{\bf G}\cdot{\bf r}}$,
we have

\begin{align}
g^{\mathrm{E}}= & \sum_{G}\widetilde{V}_{q\lambda}^{\mathrm{E}}\left({\bf G}\right)o_{kk-q}^{\uparrow\downarrow}\left({\bf G}\right),\label{eq:Elliot}\\
o_{kk-q}^{\uparrow\downarrow}\left({\bf G}\right)= & \left\langle u_{k}^{\uparrow(\downarrow)}\right|e^{i{\bf G}\cdot{\bf r}}\left|u_{k-q}^{\downarrow\left(\uparrow\right)}\right\rangle ,\label{eq:oG}
\end{align}

where $o_{kk-q}^{\uparrow\downarrow}\left({\bf G}\right)$
is ${\bf G}$-dependent spin-flip overlap function. Without loss of
generality, we suppose the first Brillouin zone is centered at $\Gamma$.

Therefore, $g^{\mathrm{E}}$ is not only determined
by the long-range component of $o_{kk-q}^{\uparrow\downarrow}\left({\bf G}\right)$,
i.e., $o_{kk-q}^{\uparrow\downarrow}\left({\bf G}=0\right)$ but also
the ${\bf G}\neq0$ components. But nevertheless, it is helpful to
investigate $o_{kk-q}^{\uparrow\downarrow}\left({\bf G}=0\right)$
and similar to Eq. \ref{eq:gsf2}, we propose an effective modulus
square of the spin-flip overlap matrix element $\overline{|o^{\uparrow\downarrow}|^{2}}$,

\begin{align}
\overline{|o^{\uparrow\downarrow}|^{2}}= & \frac{\sum_{kq}\mathrm{w}_{k,k-q}\sum_{\lambda}|o_{k,k-q}^{\uparrow\downarrow}\left({\bf G}=0\right)|^{2}}{\sum_{kq}\mathrm{w}_{k,k-q}}.\label{eq:osf2}
\end{align}

\subsection*{Internal magnetic field}

Suppose originally a system has time-reversal and
inversion symmetries, so that every two bands form a Kramers degenerate
pair. Suppose the ${\bf k}$-dependent spin matrix vectors in Bloch
basis of the Kramers degenerate pairs are ${\bf s}_{k}^{0}$ with
${\bf s}\equiv\left(s_{x},s_{y},s_{z}\right)$. The inversion symmetry
broken, possibly due to applying an electric field or a substrate,
induces ${\bf k}$-dependent Hamiltonian terms

\begin{align}
H_{k}^{\mathrm{ISB}}= & \mu_{B}g_{e}{\bf B}_{k}^{\mathrm{in}}\cdot{\bf s}_{k}^{0},\label{eq:HISB}
\end{align}

where $\mu_{B}g_{e}$ is the electron spin gyromagnetic
ratio. ${\bf B}_{k}^{\mathrm{in}}$ is the SOC field and called internal
magnetic fields. ${\bf B}^{\mathrm{in}}$ splits the degenerate pair
and polarizes the spin along its direction. The definition of ${\bf B}_{k}^{\mathrm{in}}$
is

\begin{align}
{\bf B}_{k}^{\mathrm{in}}\equiv & 2\Delta^{\mathrm{SOC}}_{k}{\bf S}_{k}^{\mathrm{exp}}/\left(\mu_{B}g_{e}\right),\label{eq:Bin}
\end{align}

where ${\bf S}^{\mathrm{exp}}\equiv\left(S_{x}^{\mathrm{exp}},S_{y}^{\mathrm{exp}},S_{z}^{\mathrm{exp}}\right)$
with $S_{i}^{\mathrm{exp}}$ being spin expectation value along direction
$i$ and is the diagonal element of $s_{i}$. $\Delta^{\mathrm{SOC}}$ is the band
splitting energy by SOC.

\subsection*{Spin-flip angle $\theta^{\uparrow\downarrow}$ for
intervalley spin relaxation}

Suppose (i) the inversion symmetry broken induces
${\bf B}_{k}^{\mathrm{in}}$ (Eq. \ref{eq:Bin}) for a Kramers degenerate
pair; (i) there are two valleys centered at wavevectors ${\bf Q}$
and $-{\bf Q}$ and (iii) there are two wavevectors ${\bf k}_{1}$
and ${\bf k}_{2}$ near ${\bf Q}$ and $-{\bf Q}$ respectively. Due
to time-reversal symmetry, the directions of ${\bf B}_{k_{1}}^{\mathrm{in}}$
and ${\bf B}_{k_{2}}^{\mathrm{in}}$ are almost opposite.

Define the spin-flip angle $\theta_{k_{1}k_{2}}^{\uparrow\downarrow}$
as the angle between $-{\bf B}_{k_{1}}^{\mathrm{in}}$ and ${\bf B}_{k_{2}}^{\mathrm{in}}$,
which is also the angle between $-{\bf S}_{k_{1}}^{\mathrm{exp}}$
and ${\bf S}_{k_{2}}^{\mathrm{exp}}$. We will prove that for a general
operator $\widehat{A}$,

\begin{align}
\left|A_{k_{1}k_{2}}^{\uparrow\downarrow}\right|^{2}\approx & \mathrm{sin}^{2}\left(\theta_{k_{1}k_{2}}^{\uparrow\downarrow}/2\right)\left|A_{k_{1}k_{2}}^{\downarrow\downarrow}\right|^{2},\label{eq:spin-filp-relation}
\end{align}

where $A_{k_{1}k_{2}}^{\uparrow\downarrow}$ and
$A_{k_{1}k_{2}}^{\downarrow\downarrow}$ are the spin-flip and spin-conserving
matrix elements between ${\bf k}_{1}$ and ${\bf k}_{2}$ respectively.

The derivation uses the first-order perturbation
theory and has three steps:

Step 1: The 2$\times$2 matrix of operator $\widehat{A}$
between ${\bf k}_{1}$ and ${\bf k}_{2}$ of two Kramers degenerate
bands is $A_{k_{1}k_{2}}^{0}$. According to Ref. \citenum{yafet1963g},
with time-reversal symmetry, the spin-flip matrix element of the
same band between ${\bf k}$ and $-{\bf k}$ is exactly zero, therefore,
the spin-flip matrix elements of $A_{k_{1}k_{2}}^{0}$ are zero at lowest order as ${\bf k}_{1}+{\bf k}_{2}\approx0$, i.e., $A_{k_{1}k_{2}}^{0,\uparrow\downarrow}\approx A_{k_{1}k_{2}}^{0,\downarrow\uparrow}\approx0$.

Step 2: The inversion symmetry broken induces ${\bf B}_{k}^{\mathrm{in}}$
and the perturbed Hamiltonian $H_{k}^{\mathrm{ISB}}$ (Eq. \ref{eq:HISB}).
The new eigenvectors $U_{k}$ are obtained based on the first-order
perturbation theory.

Step 3: The new matrix is $A_{k_{1}k_{2}}=U_{k_{1}}^{\dagger}A_{k_{1}k_{2}}^{0}U_{k_{2}}$.
Thus the spin-flip matrix elements $A_{k_{1}k_{2}}^{\uparrow\downarrow}$
with the inversion symmetry broken are obtained.

We present the detailed derivation in SI Sec. III.

From Eq. \ref{eq:spin-filp-relation}, for the intervalley
e-ph matrix elements of ML-Ge systems, we have

\begin{align}
\left|g_{k_{1}k_{2}}^{\uparrow\downarrow}\right|^{2}\approx & \mathrm{sin}^{2}\left(\theta_{k_{1}k_{2}}^{\uparrow\downarrow}/2\right)\left|g_{k_{1}k_{2}}^{\downarrow\downarrow}\right|^{2}.
\end{align}

As $\left|g_{k_{1}k_{2}}^{\uparrow\downarrow}\right|^{2}$
largely determines $\tau_{s}$ of ML-Ge systems, the differences of
$\tau_{s}$ of ML-Ge on different substrates should be mainly due
to the difference of $\mathrm{sin}^{2}\left(\theta_{k_{1}k_{2}}^{\uparrow\downarrow}/2\right)$.

For the intervalley overlap matrix elements, we should
have $\left|o_{k_{1}k_{2}}^{\uparrow\downarrow}\right|^{2}\approx\mathrm{sin}^{2}\left(\theta_{k_{1}k_{2}}^{\uparrow\downarrow}/2\right)\left|o_{k_{1}k_{2}}^{\downarrow\downarrow}\right|^{2}$.
Since $\left|o_{k_{1}k_{2}}^{\downarrow\downarrow}\right|^{2}$ is
of order 1, $\left|o_{k_{1}k_{2}}^{\uparrow\downarrow}\right|^{2}$
is expected proportional to $\mathrm{sin}^{2}\left(\theta_{k_{1}k_{2}}^{\uparrow\downarrow}/2\right)$
and have the same order of magnitude as $\mathrm{sin}^{2}\left(\theta_{k_{1}k_{2}}^{\uparrow\downarrow}/2\right)$.

Finally, similar to Eq. \ref{eq:gsf2}, we propose
an effective modulus square of $\mathrm{sin}^{2}\left(\theta_{k_{1}k_{2}}^{\uparrow\downarrow}/2\right)$,

\begin{align}
\overline{\mathrm{sin}^{2}\left(\theta^{\uparrow\downarrow}/2\right)}= & \frac{\sum_{kq}\mathrm{w}_{k,k-q}\mathrm{sin}^{2}\left(\theta_{k,k-q}^{\uparrow\downarrow}/2\right)}{\sum_{kq}\mathrm{w}_{k,k-q}}.\label{eq:sin2}
\end{align}

\section*{DATA AVAILABILITY}

The data that support the findings of this study are available upon
request to the corresponding author.

\section*{CODE AVAILABILITY}

The codes that were used in this study are available upon request
to the corresponding author.

\section*{Acknowledgements}

We thank Ravishankar Sundararaman for helpful discussions. This work
is supported by the Air Force Office of Scientific Research under
AFOSR Award No. FA9550-YR-1-XYZQ and National Science Foundation under
grant No. DMR-1956015. This research used resources of the Center
for Functional Nanomaterials, which is a US DOE Office of Science
Facility, and the Scientific Data and Computing center, a component
of the Computational Science Initiative, at Brookhaven National Laboratory
under Contract No. DE-SC0012704, the lux supercomputer at UC Santa
Cruz, funded by NSF MRI grant AST 1828315, the National Energy Research
Scientific Computing Center (NERSC) a U.S. Department of Energy Office
of Science User Facility operated under Contract No. DE-AC02-05CH11231,
and the Extreme Science and Engineering Discovery Environment (XSEDE)
which is supported by National Science Foundation Grant No. ACI-1548562
\citep{xsede}.

\section*{Author contributions}

J.X. performed the first-principles calculations. J.X. and Y.P. analyzed
the results. J.X. and Y.P. designed all aspects of the study. J.X.
and Y.P. wrote the manuscript.

\section*{ADDITIONAL INFORMATION}

\textbf{Supplementary Information} accompanies the paper on the npj
Computational Materials website.

\textbf{Competing interests:} The authors declare no competing interests.

\section*{REFERENCES}


%

\end{document}